%% file: paper.tex
\documentclass[preprint2]{aastex}
\usepackage{longtable}

\newcommand{\myemail}{sdzib@mpifr-bonn.mpg.de}

\newcommand{\rahms}[4]{$#1^{\rm h}#2^{\rm m}#3\mbox{$^{\rm s}\mskip-7.6mu.\,$}#4$} 
\newcommand{\decdms}[4]{$#1^{\circ}#2'#3\mbox{$''\mskip-7.6mu.\,$}#4$} 

\newcommand{\sbeamp}[5]{{$#1\mbox{$''\mskip-7.6mu.\,$}#2$} $\times$ {$#3\mbox{$''\mskip-7.6mu.\,$}#4$}; $+#5^\circ$}

\slugcomment{Second Draft}

\shorttitle{Deep Radio Images of the HH~124~IRS Cluster} 
\shortauthors{Dzib et al.}

\begin{document}
\newpage

\title{Deep VLA Images of the HH~124~IRS Radio Cluster and its
Surroundings and a New Determination of the Distance to NGC 2264}

\author{Sergio Dzib\altaffilmark{1,2}, 
Laurent Loinard\altaffilmark{1}, 
Luis F.\ Rodr\'{\i}guez\altaffilmark{1,3} and
Phillip Galli\altaffilmark{4} 
}

\altaffiltext{1}{Centro de Radioastronom\'{\i}a y Astrof\'{\i}sica, Universidad
Nacional Aut\'onoma de M\'exico\\ Apartado Postal 3-72, 58090,
Morelia, Michoac\'an, Mexico (\myemail)}

\altaffiltext{2}{Max Planck Institut f\"ur Radioastronomie, Auf del H\"ugel 69, 53121 Bonn, Germany}

\altaffiltext{3}{King Abdulaziz University, P.O. Box 80203, Jeddah 21589, Saudi Arabia}

\altaffiltext{4}{Instituto de Astronomia, Geof\'{\i}sica e Ci\^{e}ncias Atmosf\'{e}ricas,
Universidade de S\~{a}o Paulo, Rua do Mat\~{a}o 1226, Cidade Universit\'{a}ria, 05508-900,
S\~{a}o Paulo, SP, Brazil}

\begin{abstract}
We present new deep ($\sigma\sim6~\mu$Jy) radio images of the HH~124~IRS
Radio Cluster at 4.8 and 7.5 GHz. We detect a total of 50 radio sources, most of 
them compact. Variability and spectral indices were analyzed in order to determine 
the nature of the sources and of their radio emission. A proper motion study was also 
performed for several of these radio sources using radio observation previously
reported. Our analysis shows 
that 11 radio sources can be related with Galactic objects, most of them probably 
young stars. Interestingly, eight of these sources are in an area less than 1 square
arcminute in size. The importance of such compact clusters resides in that all its members 
can be observed in a single pointing with most telescopes, and are, therefore, ideal 
for multi-wavelength studies of variability. Another 4 of the detected sources are 
clearly extragalactic. Finally, we propose from statistical arguments that from the 
remaining sources, about 10 are Galactic, but our study does not allow us to identify
which of the sources fall in that specific category.
The relatively large proper motions observed for the sources in HH~124~IRS 
suggest that this region is located at about 400 pc from the Sun. This is
significantly smaller than the $\sim$800--900 pc distance usually assigned to the 
nearby open cluster NGC~2264 with which HH~124 is thought to be associated. 
However, a reanalysis of the Hipparcos parallaxes for members of NGC~2264, 
a convergent point approach, and a kinematic analysis all argue in favor of a
distance of order 400 pc for NGC~2264 as well. 
\end{abstract}


\section{Introduction}

Star forming regions often present large extinctions
of A$_V\sim$100 and higher. Thus, some young stellar objects (YSOs) are still
hidden in their parental clouds and cannot be detected at optical
or even near infrared wavelengths. The level of extinction
diminishes for longer wavelengths, being essentially zero in the radio 
band. Fortunately,  around 10-20\% of the YSOs detected in X-ray
and infrared studies are radio continuum emitters (e.g.\ Andr\'e 1996,
Dzib et al.\ 2013). This radio emission can have a thermal (free-free) 
origin, tracing partially ionized circumstellar material usually associated
with outflowing material (Rodr\'{\i}guez 1997). Alternatively, it can be of
non-thermal (gyrosynchrotron) origin, produced in the magnetically-active
coronae of some young stars (e.g.\ Andr\'e et al.\ 1992).
These two types of radiations can be differentiated through their observational
characteristics (see Dzib et al.\ 2013 for a recent discussion). Thermal free-free 
radiation is usually steady or only slowly variable (with timescales of years,
e.g. Rodr\'{\i}guez 1994 and Rodr\'{\i}guez et al.\ 2012), 
with a spectral index $\,\gtrsim\,-$0.1, and with no polarization.
In contrast, gyrosynchrotron radiation can show variation on timescales
of hours to days and spectral indices ranging between very negative
values in the optically thin case and $+2.5$ in the optically thick limit.
It can also show significant levels of circular polarization 
(e.g. G\'omez et al.\ 2008). 

During a program of studies of compact radio emission
from regions of star formation, Reipurth et al.\ (2002)
detected a remarkable compact ($\sim 30''$) cluster of six radio objects 
associated with the HH~124~IRS region. Their flux densities
were around 0.1 mJy on the average. All six sources
(named VLA 1, 2, 7, 8, 9, and 10 by Reipurth et al.) appear to be
variable on a timescale of days, showing excursions
of several times 0.01 mJy from one day to the
next. However, since the rms noise of the observations
of Reipurth et al.\ (2002)
for a given day were of order 0.01 to 0.02 mJy, 
comparable with the apparent variations,
new observations of very high sensitivity were
needed to test the reality of the variations.
According to Reipurth et al.\ (2002), in the area where the six
sources were found, less than one extragalactic source
was expected given the level of sensitivity reached. Finally, since three
of these radio sources have near-IR (2MASS) or optical
counterparts, they were considered most
likely young stellar objects (YSOs).

Here we report new, high sensitivity radio observations, 
performed with the Karl G. Jansky Very Large Array (VLA),
of the HH~124~IRS region. Our main objective was to test 
the reality of the variations in the emission of
the components of the compact radio cluster
and to determine the nature of their emission. Also,
an astrometric study was performed by combining our observations
with those of Reipurth et al.\ (2002),
in order to search for proper motions of the sources
in the cluster.

\section{Observations}

The observations were made with the VLA of the NRAO\footnote{The 
National Radio Astronomy Observatory is operated 
by Associated Universities
Inc. under cooperative agreement with the National Science Foundation.}
at C band (4 -- 8 GHz) in the B configuration. 
The observations were split in two sub-bands, each with a
bandwidth of 1 GHz, centered at 4.8 and 7.5 GHz. 
A total of four epochs were obtained.
The first epoch was on 2012 June 9, and the remaining three were
on three consecutive days from 2012 August 8 to 10.

At the beginning of each observation, we observed the standard flux calibrator 
3C 138 for $\sim$10 minutes. Then we spent one minute on the phase calibrator
J0632+1022 followed by nine minutes on the target; this cycle was repeated until
the two hours of observation at each epoch were completed. 
The angular distance between the phase calibrator and the target is 2.54 degrees.
{The target phase center is RA = \rahms{06}{41}{02}{841} and
Dec. = \decdms{+10}{14}{58}{32},
it was chosen to be the center of the compact cluster
reported by Reipurth et al.\ (2002).}

The data were edited, calibrated and imaged in the standard fashion using the 
Common Astronomy Software Applications package (CASA\footnote{CASA is 
developed by an international consortium of scientists based 
at the National Radio Astronomical Observatory (NRAO), the European 
Southern Observatory (ESO), the National Astronomical Observatory of 
Japan (NAOJ), the CSIRO Australia Telescope National Facility (CSIRO/ATNF), 
and the Netherlands Institute for Radio Astronomy (ASTRON) under the 
guidance of NRAO.}). After the calibration, the visibilities
were imaged with a pixel size of 0.14 and 0.20 arcsecond, for the 7.5 and 4.8
GHz sub-bands, respectively. The weighting scheme used was intermediate between
natural and uniform (WEIGHTING='briggs' with ROBUST=0.0 in CASA). 
The minimum level of the primary beam used was at
a response of 20\%. The final images covered
circular areas of 10 and 16 arcminutes in diameter, for the 7.5 and 4.8 
GHz sub-bands, respectively, and were corrected for the effects of the
position-dependent primary beam response.
The rms noise levels in the central part
of the final images at each epoch
were around 5.5 and 6.0 $\mu$Jy beam$^{-1}$, in the 7.5 and 4.8 GHz 
sub-bands, respectively.
To obtain deeper images, we concatenated the four epochs,
and then imaged the visibilities using the same scheme described above.
The rms noise levels in the central part of these final images were 
2.8 and 3.2 $\mu$Jy beam$^{-1}$, for the 7.5 and 4.8
GHz sub-bands, respectively. { The synthesised beams at 7.5 GHz
sub-band are
(\sbeamp{0}{72}{0}{66}{26}), 
(\sbeamp{0}{79}{0}{71}{111}), 
(\sbeamp{0}{74}{0}{70}{131}), and
(\sbeamp{0}{79}{0}{69}{122}) from
the first to the last epoch, respectively. In the same way for the
4.8 GHz sub-bands are 
(\sbeamp{1}{14}{1}{01}{17}), 
(\sbeamp{1}{19}{1}{07}{123}), 
(\sbeamp{1}{13}{1}{05}{141}), and
(\sbeamp{1}{21}{1}{05}{129}).
For the images with
the four concatenated epochs, the synthesised beams are
(\sbeamp{0}{73}{0}{69}{146}) and (\sbeamp{1}{12}{1}{04}{155}), for 
the 7.5 and 4.8 GHz sub-bands, respectively.
Using these images, we determined the source positions and flux
densities using a two-dimensional fitting procedure
(task IMFIT in CASA).}

\section{Results}

We first searched for sources in the deep images obtained
by concatenating all four epochs.
We considered a detection reliable
if the source was above three times the rms noise level in the localized area
and had a previously known counterpart. If there were no known counterparts,
then five times the rms noise level was required in at least one of the sub-bands.
This procedure was adopted to minimize the possibility of reporting a large
noise fluctuation as a real source.

With these requirements, a total of 50 sources were detected (Table 1). 
Of these, 28 were detected in both sub-bands, while 20 were detected 
only in the 4.8 GHz sub-band and 2 only in the 7.5 GHz sub-band. However, 
13 of the 20 sources detected only in the 4.8 GHz sub-band fall outside the
field of view of the 7.5 GHz image, so only 7 were undetected at 4.8
GHz because their fluxes are below three times the rms noise level.
The positions
and flux densities of each source
from these images are listed in Table 1.
{ Two sources of uncertainties on the fluxes are included: (i) 
the error that results from the statistical noise in the images, and (ii) a systematic
uncertainty of 5\% resulting from possible errors in the absolute flux calibration.}
 The spectral index 
{$\alpha$, where S$_{\nu}\propto\nu^{\alpha}$,} was determined for those sources detected in both
sub-bands, and are also listed in Table 1. After the detection in the deep images,
the flux density and position of each source was measured in 
the images corresponding to the 
individual epochs. For these images, we considered a non detection when the
source was below three times the rms noise level.

Once the flux densities had been obtained from each epoch, they were
compared to check for variability. The percentage of variability
was measured as the ratio between the largest flux difference 
(maximum minus minimum flux densities) 
and the maximum flux density. We did not search
for variability in sources
with extended emission, because sensitivity and uv coverage
effects can produce spurious variations.
In addition, we included the uncertainty in the variability that
is introduced by absolute pointing errors of the primary beam of
the VLA antennas, as described in the Appendix.
The percentage of variation in
each sub-band is listed in Table 1. This variation was estimated using the
four epochs for the sources that were detected
at more than one epoch and are not extended. We considered as statistically significant only those
variations that are above 3-$\sigma$. From Table 1 we can see that 
10 sources (out of a total of 43 where variability could be measured)
are variable at 4.8 GHz, while 3 (out of
a total of 23) are variable at 7.5 GHz. If we restrict the analysis
to the three consecutive epochs, we find that only 4 of the sources at 4.8 GHz are
variable, while none of the 7.5 GHz sources is variable. Sources with such short-term 
variability are indicated in the table. The largest variations observed here are of
the order of a factor of two.

\subsection{Counterparts of the detected radio sources}

Given the position of the sources, we looked for counterparts to the
detected radio sources using the {\it SIMBAD astronomical database}. 
A source from {\it SIMBAD} was considered a counterpart if it was separated
from a radio source reported here by less than
1.0 arcsec. We detected 8 of the 10 sources reported by Reipurth
et al.\ (2002); the two exceptions being VLA 3 and 8. VLA 3 was clearly detected
by Rodr{\'{\i}}guez \& Reipurth (1998) with a flux density of 0.60 mJy at
8.3 GHz
and considered by them to be a background source. We do not detect it
at a level of $\sim$11 $\mu$Jy beam$^{-1}$, that represents three times the
rms noise, at each sub-band, in the area where it was previously reported.
This difference in flux densities suggests a large variability, of order $\sim$50.
However, the observations of Rodr{\'{\i}}guez \& Reipurth (1998) were made
with lower angular resolution ($\sim9{''}$) than ours and we may be simply resolving out the
source.
VLA 8, also known as LkHA 46, is a YSO that was first detected in the radio by Reipurth
et al.\ (2002), and showed strong flux variability, by a factor larger
than 2, on timescales of days.
We do not detect it here at three times the rms noise of 8.4 and 9.6 $\mu$Jy
for 7.5 and 4.8 GHz sub-bands, respectively. From these results, 
we conclude that LkHA 46 is also strongly variable on timescales of years. 

\begin{figure*}
\begin{center}
\begin{tabular}{l}
\includegraphics[width=00.50\linewidth,trim= 50 0 0 0]{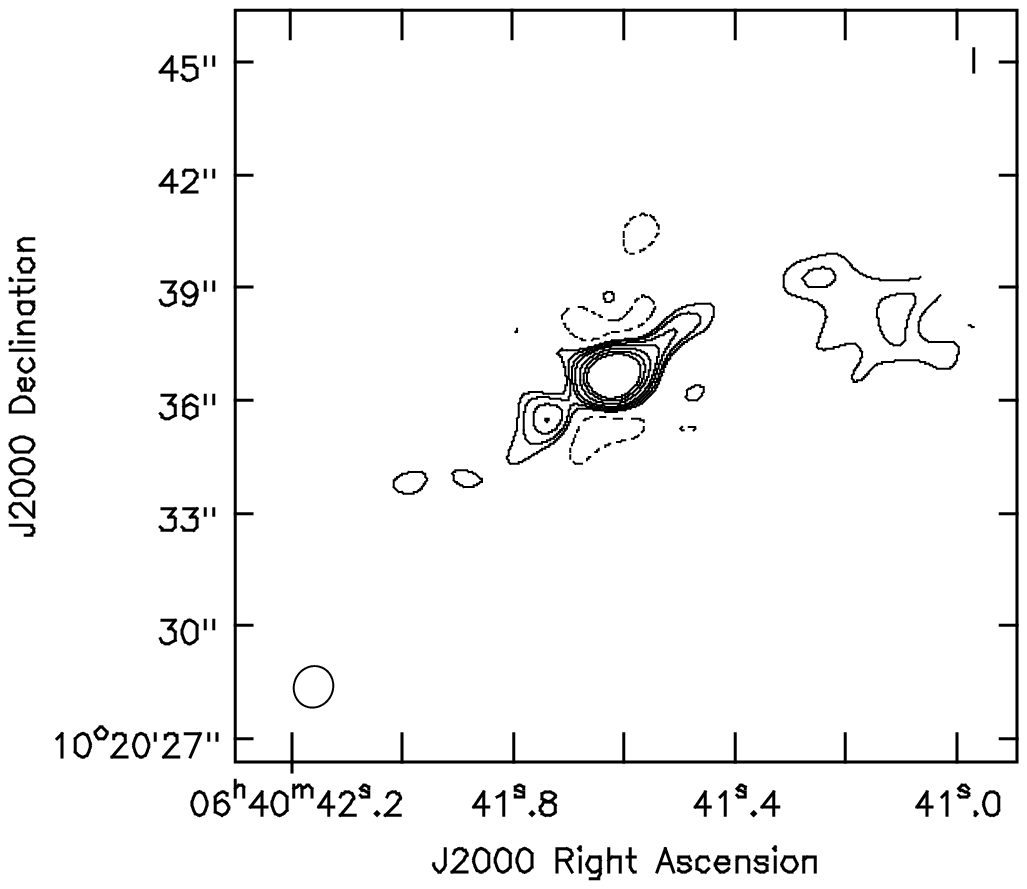}\\
\includegraphics[width=00.50\linewidth,trim= 55 0 0 0]{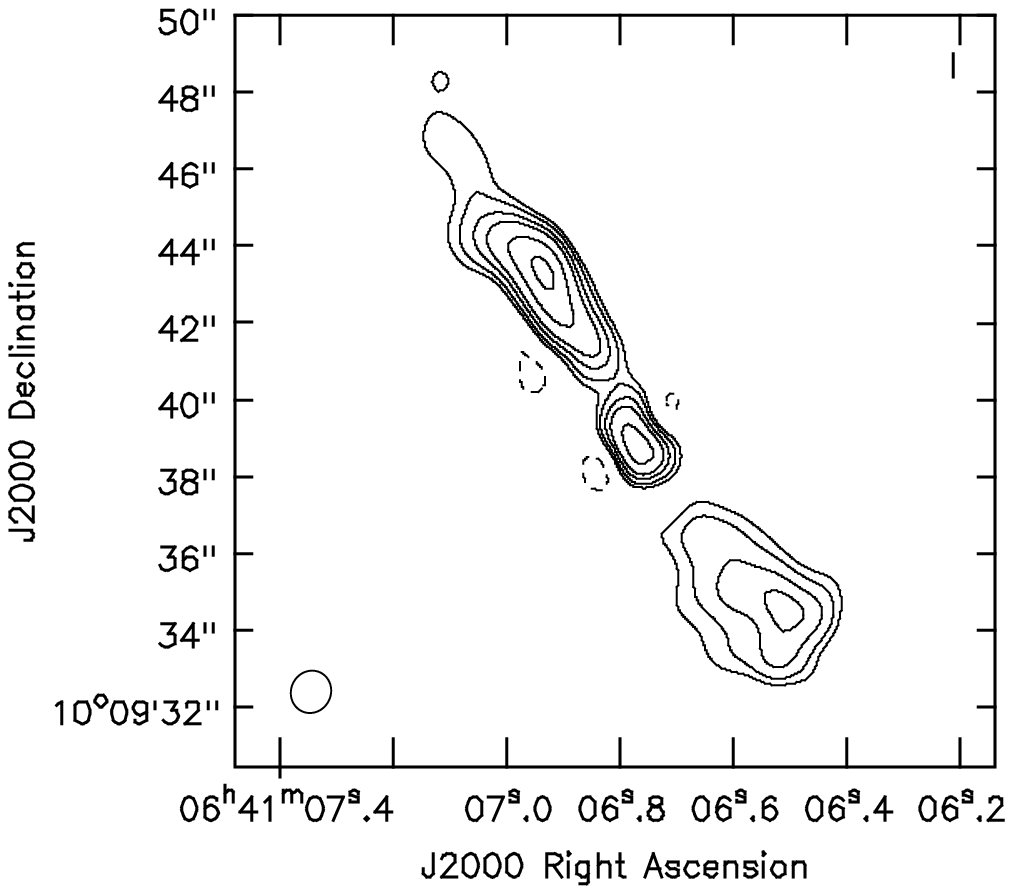}\\
\includegraphics[width=00.50\linewidth,trim= 58 0 0 0]{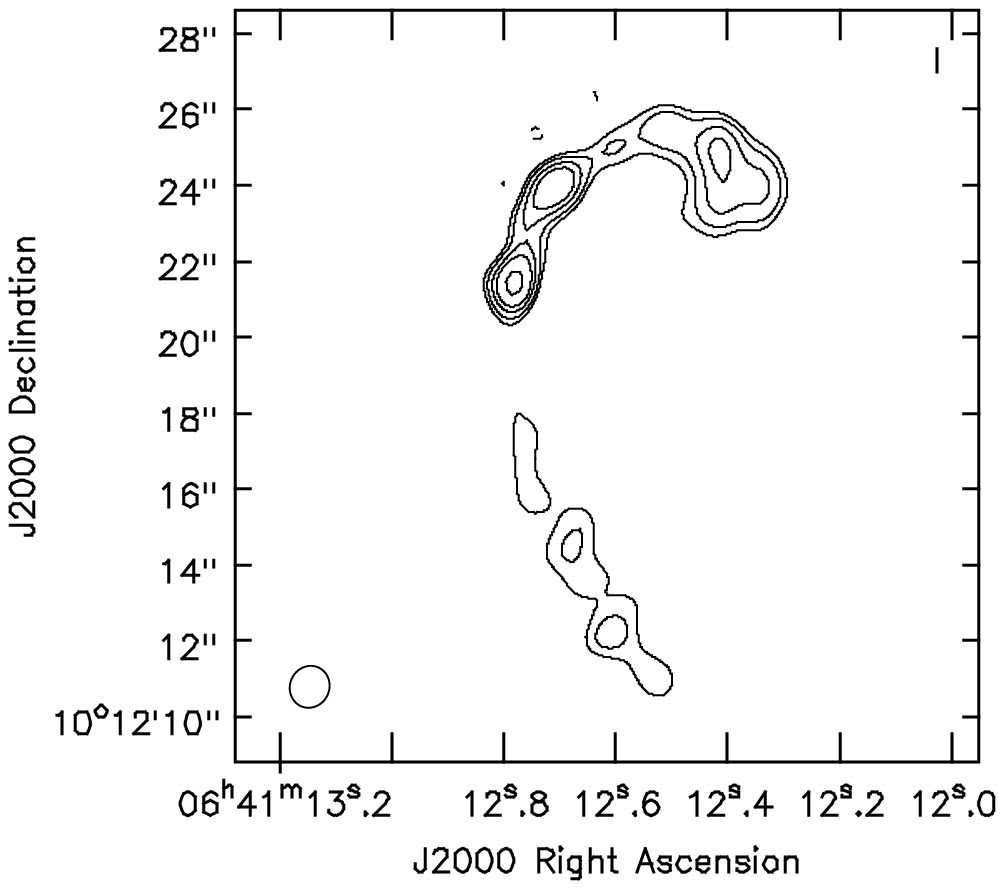}
\end{tabular}
\end{center}
\caption{ VLA contour images, from the four concatenated epochs,
of the 4.5 GHz emission of the HH124-VLA J064041.61+102036.6 (top),
HH124-VLA J064106.77+100938.8 (center) and HH124-VLA J064112.78+101219.1 (bottom) radio sources. 
Contours are -4, 4, 6, 9, 12, 18, 24 and 30 times 
the rms noise in the localized area, 31  $\mu$Jy beam$^{-1}$, 16 $\mu$Jy beam$^{-1}$
and 7.3 $\mu$Jy beam$^{-1}$ for HH124-VLA J064041.61+102036.6,
HH124-VLA J064106.77+100938.8 and HH124-VLA J064112.78+101219.1, respectively.
Together with VLA 6, these are the only extended sources detected in our observations.}
   \label{fig:extended}
\end{figure*}

The 2MASS, WISE and SDSS8 catalogs (Cutri et al.\ 2003, Cutri et al.\ 2012
and Adelmann-McCarthy\ 2011, respectively) were also checked and some
counterparts  to the radio sources were found (see Table 2). HH124-VLA J064102.59 +101503.6 = VLA 1 and
HH124-VLA J064102.65 +101501.8 = VLA 9, have infrared counterparts and are
known to be YSOs. Together with VLA 8, they are the only radio sources
confirmed to be YSOs. Another two of them, HH124-VLA J064041.61+102036.6 {(shown in
Figure} \ref{fig:extended}) and 
HH124-VLA J064104.33+101240.9, are classified as extragalactic objects in 
the SDSS8. The object HH124-VLA J064118.28+101745.7 was associated to a stellar object
in the SDSS8. The remaining sources have not been related to any kind of object but
their low variability in the radio (Table 1) resembles that of extragalactic objects.
Finally, in the cases of HH124-VLA J064106.77+100938.8, HH124-VLA J064112.78+101219.1
and HH124-VLA J064114.16+101158.5 = VLA 6, { additionally to their low variability,}
their jet like morphologies also suggest an extragalactic nature { (see Figures
\ref{fig:extended} and \ref{fig:vla6}), but we cannot discard a Galactic nature}.

\subsection{Proper motions}

\begin{figure*}
\begin{center}
\includegraphics[width=01.00\linewidth,trim= 100 550 180 80]{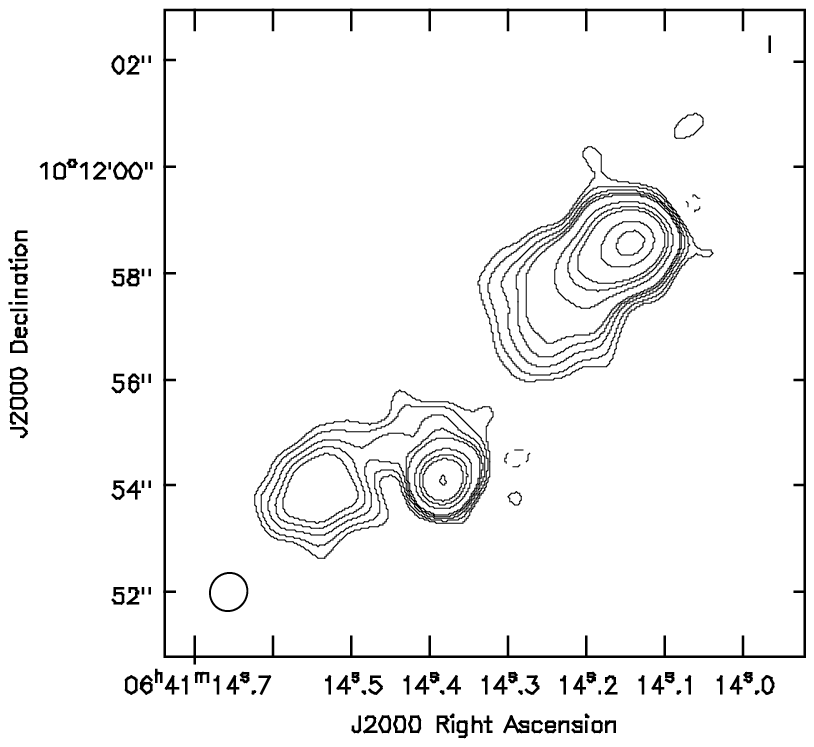}
\end{center}
\caption{VLA contour image, from the four concatenated epochs,
of the 7.5 GHz emission of the HH~124~IRS VLA 6 radio source. 
Contours are -4, 4, 6, 9, 12, 15, 30, 45, 60, 120 and 180 times 
the rms noise in the localized area, 18  $\mu$Jy beam$^{-1}$. Three
components are resolved. The northernmost one is also the strongest
and the one reported by Reipurth et al.\ (2002).}
   \label{fig:vla6}
\end{figure*}

In order to investigate more about the nature of the radio sources
around HH~124~IRS, we used the archive data reported by Reipurth et al.\ (2002)
to measure the proper motions, $\mu$, of the compact bright objects detected in both 
sessions. A total of eight objects fulfill these requirements and are listed in Table 3.
We also include VLA 6 although it is resolved into three different components
(see Figure \ref{fig:vla6}). In this case, we consider only the northernmost component, which 
is the strongest, and use the position reported by Reipurth et al.\ (2002) corresponding 
to it. This component is resolved, and we obtain the position just to the brightest region.  
Its central component was marginally detected and resolved by Reipurth et al.\ (2002),
and we do not  perform astrometry with it. The third component was not detected by Reipurth et al.\ (2002).

The data from Reipurth et al.\ (2002) were constituted by four 11-hour observations,
taken on four consecutive days, from 2000 November 26 to 29, at 8.4 GHz. 
These data were edited and calibrated following
the standard VLA procedures and using the software package AIPS. These observations
used the quasar J0613+131 as the phase calibrator. Even though the phase calibrator
is not the same for both sessions, their positions are well known and reliable 
astrometry can be performed. The position of quasar J0632+1022 is known to a
level of 10 mas, that it is lower than the statistical errors in position obtained for C-band
observations in B configuration. In addition, it is closer to the studied region
and stronger than J0613+131, so the phase transfer is better. 

In Table 3, the proper motions in Right Ascension, $\mu_\alpha\cos{\delta}$, and Declination
$\mu_{\delta}$, are presented. The total proper motions, $\mu_{\rm total}$, of most
sources are consistent with a motion of $\sim$ 7 mas yr$^{-1}$, except for VLA 6.
VLA 6 was suggested to be an extragalactic object by Reipurth et al.\ (2002),
so proper motions are not expected from this source. Indeed, its measured proper
motion is consistent with no movement at the 1 $\sigma$ level. We note that the lack of detectable
motion for this source confirms the reliability of our astrometry.

\section{Discussion}

\subsection{Kinematics of the region and individual objects}

As already noted, most proper motions follow the tendency of an absolute value of
$\mu_{\rm total}\sim$ 7 mas yr$^{-1}$ in the southeast direction. 
The different phase calibrators used in both sessions could produce 
this apparent effect. However, the negligible proper motions of VLA 6,
shows that this effect is not important. A more reasonable explanation is that we are measuring 
the intrinsic motion between the star forming region and the Sun. As YSOs in a star forming region 
are formed from the same cloud, it is expected that the young stars conserve the 
memory of the movement of the cloud and, thus, that they will share similar proper motions. 
We favor this effect as the main contributor to the proper motion tendency.
In what follows we discuss some of the sources with measured proper motions
in more details.

{\bf HH124-VLA J064101.22+101300.5 = VLA 7}. Its total proper motion is the largest
in our sample and its direction follows that of the group.
This is consistent with it being a YSO.

{\bf HH124-VLA J064110.65+101538.2 = VLA 4}. This source was previously suggested to be an
extragalactic object (Reipurth et al.\ 2002).
However, it has a positive spectral index and it is slightly variable.
Its proper motion is the second largest in our sample and is in a similar 
direction as that of the other sources in the cluster. These properties strongly suggest that it is a 
Galactic object, most probably related to the HH~124~IRS cluster. 

{\bf HH124-VLA J064112.23+101417.2 = VLA 5}. This source was also suggested to be an extragalactic
object (Reipurth et al.\ 2002). It has a flat spectral index and low variability.
Within the errors in the spectral index it is not possible to favor
if it is a Galactic or extragalactic object. Its proper motion suggests
that is a Galactic object and probably related to the compact cluster HH~124~IRS.

{\bf HH124-VLA J064056.62+101408.7}. The negative spectral index and the low
variability of this source suggest an extragalactic origin. Its proper motions
are consistent with no motion within two times the errors. Interestingly,
however, the measured proper motions are similar to the estimated average of 
the other considered YSOs. As a consequence, we do not discard a Galactic nature.

\subsection{Counts of Background Sources}

From Table 1, the number of radio sources classified as definite or likely extragalactic sources
is 39. All of them are detected in the 4.8 GHz sub-band, while only 20 are detected
in the 7.5 GHz sub-band. Seven of these sources were not detected in the 7.5 GHz sub-band
because they fell below three times the noise level in the image. The remaining 12 were outside
of the field of view covered by the 7.5 GHz observations.

In order to determine if the number of background sources is consistent with the expected 
values, we compare them with the values given by the formulation of Anglada et al.\ (1998).
Considering a minimal detection flux of 14 $\mu$Jy and 16 $\mu$Jy at 7.5 and 4.8 GHz,
the number of expected sources are $N_{8.4}\simeq 9$ and $N_{4.8}\simeq 30$. This 
estimate suggests that we are attributing an extragalactic nature to $\sim$10 sources that 
are actually Galactic. Unfortunately, with the present data it is not possible to establish 
which sources these might be.

\subsection{The compact cluster of radio YSOs around HH~124~IRS}

\begin{figure*}
\begin{center}
\includegraphics[width=0.85\linewidth,trim= 130 570 160 0]{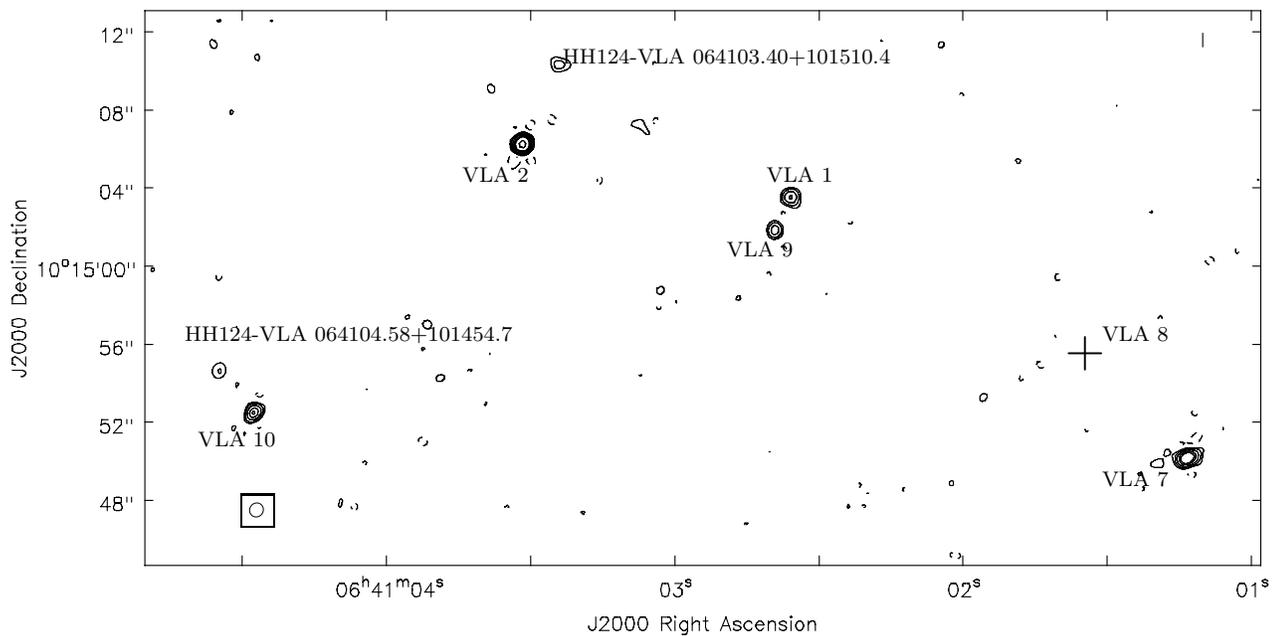}
\put(-363.50,34.50){\framebox(12,12)}
\put(-52.0,95.0){{\huge +}}
\put(-38.0,105.0){\footnotesize {VLA 8}}
\put(-38.0,50.0){{\footnotesize VLA 7}}
\put(-165.0,165.0){{\footnotesize VLA 1}}
\put(-180.0,137.0){{\footnotesize VLA 9}}
\put(-380.0,65.0){{\footnotesize VLA 10}}
\put(-385.0,105.0){{\footnotesize HH124-VLA 064104.58+101454.7}}
\put(-280.0,165.0){{\footnotesize VLA 2}}
\put(-242.0,210.0){{\footnotesize HH124-VLA 064103.40+101510.4}}
\end{center}
\caption{VLA contour image of the 7.5 GHz emission of the HH~124~IRS radio cluster. 
Contours are -3, 3, 5, 8, 12, 15, and 25 times the rms noise. This image was made by
concatenating the data from the four epochs observed here, and has an rms noise of 
2.8 $\mu$Jy beam$^{-1}$. The rms noise of the individual days is about 6 $\mu$Jy 
beam$^{-1}$. The cross indicates the position of VLA 8, which was not detected in our 
recent observations, but was detected up to 12 times the noise level in the images 
of Reipurth et al.\ (2002).}
   \label{fig:clust}
\end{figure*}

Our observations confirmed the existence of a remarkable cluster of radio sources 
concentrated to an area of less than 1 arcmin$^2$ initially reported by Reipurth et al.\ 
(2002). In addition to the original six sources reported by Reipurth et al.\ (2002), we 
detect two new objects in this area (see Figure \ref{fig:clust}). In such a compact area, 
less than one background object is expected. Reipurth et al.\ (2002) reported possible 
day-to-day variability of the order of 10 $\mu$Jy, but as the noise level was of the same 
order as the variability, it was not clear whether or not the variability was real. Our 
results discard strong day-to-day variability for most of these radio sources, as only 
the source VLA 10 showed clear short-term variations. However, it should be noted 
that 4 of the remaining 6 sources in the radio cluster have variability above 2-$\sigma$ 
(but below our threshold of 3-$\sigma$), hinting again at the presence of low level 
day-to-day variability.

In addition to the cluster around HH~124~IRS, our results show that four other objects 
have a Galactic origin. HH124-VLA J064118.28+101745.7 is classified as a star in the 
SDSS8 survey. Its spectral index is not well determined because it was not detected in 
the 7.5 GHz sub-band. Its variability seems to be low. With the existing data,  we cannot 
ascertain whether it is part of the cluster or a background/foreground object. 
HH124-VLA J064105.92+101551.7 is definitely a variable radio source, showing 
variation on timescales of days. This kind of behavior is typical of YSOs.
The other two sources, VLA 4 and 5, show a steady flux. The spectral index
of VLA 4 is positive which suggest that it is a YSO. VLA 5 shows a flat
spectrum. The direction of the proper motions for both sources suggest that they
are related to the compact cluster. 

The spectral indices for most of the radio sources in the compact cluster are positive. 
Fluxes produced by thermal processes in YSOs are usually steady and consistent with 
no variations at all (Rodr\'{\i}guez 1994). On the other hand, gyrosynchrotron emission 
from YSOs can be variable, on timescales as short as hours {(Andr\'e 1996)}. Spectral 
indices for gyrosynchrotron emission vary from very negative to $+$2.5 (Dulk 1985). 
We suggest that both mechanisms are present in the different stars in the cluster, 
with thermal emission being the dominant process.

{ In a series of papers, Rodriguez \& Reipurth (1994), Rodr{\'{\i}}guez \& Reipurth (1996),
Rodr{\'{\i}}guez \& Reipurth (1998), Reipurth et al.~(1999), Rodr{\'{\i}}guez et al.~(2000),
Reipurth et al.~(2002) and Reipurth et al.~(2004) observed a total of 28 regions of low mass
star formation. HH 124, with a cluster of 8 radio sources in a region of only $\sim 30{''}$
is the one with the densest concentration of sources per solid angle.} This kind of compact 
clusters of YSOs with radio emission are rare and very useful to test the correlation between
radio emission and emission at other wavelengths because they can be monitored 
simultaneously at radio, near-IR and  X-ray wavelengths with single pointing observations.
HH~124~IRS appears to be an ideal candidate for this kind of studies. 

{

\section{Distance to HH~124 and the nearby cluster NGC~2264}

The distance to HH 124~IRS is poorly known. This region is usually associated to the nearby
(in the plane of the sky) NGC~2264 cluster (see e.g. Reipurth et al.\ 2002), for which a distance 
of 913$\pm$117 pc has been 
recently determined (Baxter et al.\ 2009). In this section, we will use our radio determination of
the proper motions of sources in HH~124~IRS to estimate the distance to this cluster. We will 
also discuss the relation between HH~124 and NGC~2264, and revisit the distance determination
to the latter.

\subsection{A generalized kinematic distance to HH 124 IRS}

Kinematic distances are often used for Galactic objects when no
trigonometric parallaxes are available. Under this
strategy, a Galactic rotation curve is assumed, and the expected 
value of the radial velocity is calculated as a function of distance
for the line of sight corresponding to the Galactic coordinates 
$(\ell, b)$ of the object of interest. The distance 
ascribed to the object is that for which the expected value best 
matches the observed radial velocity. Evidently, a similar approach
can be followed to estimate distances from the measured components
of the proper motion. Here, the expected values of the two components 
of the proper motion will be calculated as a function of distance (assuming
a rotation curve) for the line of sight of interest, and the distance
ascribed to the source will be that for which the calculated value best 
matches the observed proper motion. Two different (but not entirely
independent) estimates of
the distance (one for each component of the proper motion) will be 
obtained by this method. Of course, if both the proper motion and the
radial velocity are known for a given source, the traditional kinematic
distance can also be measured, and three different estimates of
the distance can be obtained. Kinematic distances based on proper
motions have rarely been used because (i) accurate proper motions
are much harder to measure than accurate velocities, and (ii) 
proper motions are angular velocities (i.e.\ space velocities 
divided by distance) so they rapidly become undetectably small
with increasing distance.

Both the traditional kinematic distance method and its extension to 
proper motions are obviously subject to uncertainties. First, the observational
errors on the radial velocity and proper motion determinations will 
propagate to result in an error on the inferred distance. Second, a
rotation curve has to be assumed, and any error associated to that
choice will affect the measured distances. A third source of error comes
from the chosen value of the Solar motion required to transform velocities
from the heliocentric to the LSR rest frame. While a value $(U, V, W)$ 
of $(10.00, 5.25, 7.17)$ km s$^{-1}$ was widely used until recently
following Dehnen \& Binney (1998), a more recent determination
by Sch\"onrich et al.\ (2010) favours $(11.10, 12.24, 7.25)$ km s$^{-1}$. 
We will use this latter value in the present work. Finally, the 
kinematic distance  assumes that the objects move on circular orbits, 
as described by the rotation curve. In reality, there is also an unknown 
peculiar velocity contribution. In the case, considered here, of young 
stars recently formed out of interstellar material, this contribution can 
be assumed to be reasonably small (of order 10--15 km s$^{-1}$ 
--Binney \& Merrifield 1998). In addition, the velocity dispersions in 
radial / azimuthal / vertical Galactic directions are expected to be in the 
ratio of 1$\div$0.61$\div$0.47 (Binney \& Merrifield 1998). Here, we will
assume dispersions of 10.0 km s$^{-1}$ in the radial direction,
6.1 km s$^{-1}$ in the azimuthal direction, and 4.7 km s$^{-1}$
in the vertical direction. This corresponds to a total velocity dispersion
of 12.6 km s$^{-1}$.

The radial velocity of the source driving the most prominent outflow 
in HH~124 is $V_{lsr}$ = $+$7.2 $\pm$ 1.6 km s$^{-1}$ (Margulis et al.\
1988), corresponding to V$_{hel}$ = 19.8 $\pm$ 1.6 km s$^{-1}$. On 
the other hand, from the radio observations presented here, we can 
estimate the proper motion of the HH~124~IRS cluster by averaging
the values obtained for the seven sources identified as YSOs or candidate 
YSOs (Table 3). We obtain $\mu_\alpha\cos{\delta} = -5.0 \pm 3.1$ mas~yr$^{-1}$, 
$\mu_\delta = -5.4 \pm 2.0$ mas~yr$^{-1}$. From this dataset, we can
obtain three different kinematic estimations of the distance to HH~124~IRS
(the traditional kinematic distance based on radial velocity, and two
estimates based on the two components of the proper motion).  The
method is illustrated in Figure \ref{fig:kin}, and the results presented in
Table \ref{tab:kin}. To assess the effect of the rotation curve choice, we
have considered four options: (i) the traditional rotation curve of Brand 
\& Blitz (1993); (ii) a flat rotation curve with the IAU recommended value
of 220 km s$^{-1}$ for the LSR circular speed; (iii) the more recent
rotation curve proposed by Reid et al.\ (2009); and (iv) the flat rotation
curve also proposed by Reid et al.\ (2009) with an LSR circular speed
of 254 km s$^{-1}$. The impact of peculiar velocities is shown as "theoretical 
error bars" in Figure \ref{fig:kin} and has been taken into account in the
final distance uncertainties. Figure \ref{fig:kin} and Table \ref{tab:kin} 
shows that there is an overlap between the distance determinations 
produced by the three kinematic methods, so it is meaningful and 
advantageous to form the weighted mean of the three values. Also, it is 
clear from Table \ref{tab:kin} that the choice of rotation curve 
only has a modest effect (of order 10\%) on the final error budget. The 
errors are dominated by the observational uncertainties, and unknown
peculiar velocities. In the rest of the paper, we will use the most recent
rotation curve (by Reid et al.\ 2009), but emphasize that our results do not
strongly depend on this choice.

The distance to HH~124~IRS favoured by our kinematical analysis is
404 $\pm$ 127 pc (Note that the proper motions alone favor an even
shorter distance of order 350 pc.) This is significantly shorter that the 
distance to the nearby cluster NGC~2264 favored in the recent literature 
(913$\pm$117 pc). Yet, there is evidence for interactions between HH~124 
and NGC~2264 (see e.g. Reipurth et al.\ 2002). Let us now discuss the 
distance to NGC~2264.

\begin{figure*}
\begin{center}
\includegraphics[width=0.90\linewidth]{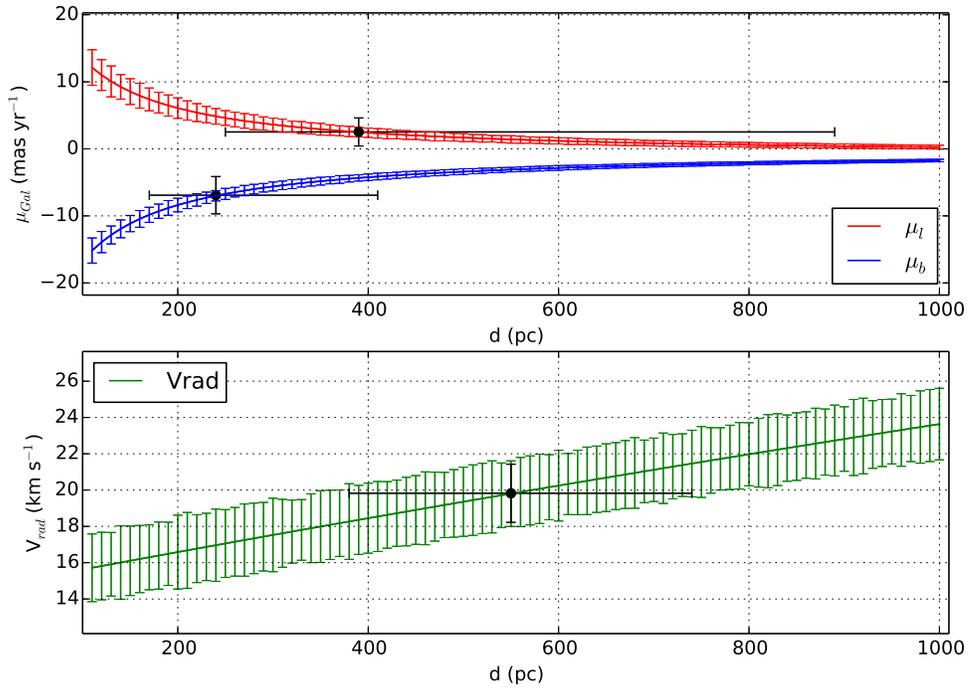} 
\end{center}
\caption{(top:) Proper motion in Galactic longitude and latitude as a function
of distance for the Galactic rotation curve of Reid et al.\ (2009). The error
bars indicate the uncertainty due to the possible peculiar velocity of the 
source. The black points show the measured proper motion for HH~124~IRS
and the error bars on that point indicate the range of possible distances.
(bottom:) Same as top panel, but for the radial velocity.}
   \label{fig:kin}
\end{figure*}

\subsection{A review of existing distance estimates for NGC~2264}

NGC~2264 is a young
open cluster related to the Mon~OB~1 association that has
been the focus of many studies of stellar evolution.
Distance estimates for NGC~2264 reported in the literature
from 1950 to 1985 range from 500~pc to 875~pc
(see Table~IX of Perez et al.\ 1987), but more recent 
determinations yield  950$\pm$75~pc (Perez et al.\ 1987),
910$\pm$50 (Neri et al.\ 1993) and 760$\pm$85 (Sung et 
al.\ 1997). Most of these distance determinations have
been derived by the same technique, i.e. cluster fitting
of the HR-diagram, and the different results reflect different 
choices of sample of stars, photometry, 
and cluster reddening by different authors. In a recent paper,
Baxteret al.\ (2009) employed a different technique and
derived a distance of 913$\pm$40~pc to NGC~2264 by using
a method that takes the projected rotational velocities
of cluster members and models the observed distribution
of $\sin i$ (where $i$ is the inclination angle of the
stellar rotational axis). 


As a first approach to examine further the distance to NGC~2264, 
we searched the new version of the Hipparcos catalog (van
Leeuwen 2007) for stars in NGC~2264 with known
parallax. We found trigonometric parallaxes for five
`classical' cluster members (S Mon, HD~47777, HD~47961,
HD~48055 and HD~47732) identified by Walker (1956).
Recently, van Leeuwen (2009) revisited the properties
of the NGC~2264 cluster and included in his analysis
another four stars (HD~47662, HD~47751, HIP~32141 and
HIP~32245) in the cluster field that exhibit proper motions
and parallax values consistent with the `classical' members
listed above. These stars are given in Table~\ref{tabX}
together with their proper motions and radial velocities
(to be discussed later in this section). The distances
corresponding to these trigonometric parallaxes 
ranges form 254 to 970 pc. Interestingly, the
trigonometrical parallax for S Mon, the dominant
stellar component of this cluster, is $\pi=3.55\pm0.50$~mas
corresponding to a distance of  $281^{+47}_{-34}$~pc. The two analysis performed by
van Leeuwen (2009) for the NGC~2264 cluster gives parallaxes
of $2.30\pm0.46$~mas (435$^{+109}_{-72}$ pc) and
$1.81\pm0.40$~mas (552$^{+157}_{-100}$ pc). 
The first was using the data for all nine
stars and the second was obtained by excluding (somewhat
arbitrarily) the three
stars with parallax near 3.5~mas, i.e. distances shorter
than 300 pc. The weighted mean
of the nine parallaxes listed in Table~\ref{tabX} is
$\pi=2.63\pm0.27$~mas ($d=380_{-36}^{+44}$~pc);  if we
retain only the five `classical' cluster members given in
the upper panel of Table~\ref{tabX} we find instead 
$\pi=2.66\pm0.33$~mas ($d=376_{-41}^{+53}$~pc) showing that
both results are consistent between themselves. In summary,
all Hipparcos estimates of the distance to the NGC~2264 cluster 
lie between 280 and 550 pc, with a mean around 405 pc. This
is inconsistent with the indirect methods that we mentioned
earlier, but in excellent agreement with our generalized kinematic method
result for HH~124~IRS.


\subsection{A convergent point distance to NGC 2264}

To make further progress, we queried the CDS database for stars 
and YSOs related to the NGC~2264 cluster within $30\arcmin$ of its
central position, $\alpha=06^{h}40^{m}58^{s}$,
$\delta=09^{\circ}53.7\arcmin$ (Wu et al.\ 2009), and
found 4004 stars previously identified in many studies
(Walker 1956; Sagar \& Joshi 1983; Sung et al.\ 1997;
Flaccomio et al.\ 1999, 2000, 2006; Rebull et al.\ 2002;
Sung et al.\ 2004; Lamm et al.\ 2004; Ram\'{\i}rez et
al.\ 2004; Dahm \& Simon\ 2005; Dahm et al.\ 2007). We
searched for proper motion data in the Tycho-2 (H{\o}g 
et al.\ 2002), UCAC4 (Zacharias et al.\ 2013) and PPMXL
(Roeser et al.\ 2010) catalogs, and found at least one
proper motion measurement for 2024~stars. Among the
various proper motions measurements we select the ones
with the smallest uncertainties. We reject all stars
with proper motion values that are not significant
(i.e., $\sigma_{\mu}>\mu$), because of measurement errors,
and stars that exhibit discrepant proper motion data
(after a $3\sigma$ elimination). This leaves us with
584~stars in the sample. We found radial velocity
information for only 58~stars in that sample. The main
source of radial velocities is F{\H u}r{\'e}sz et
al.\ (2006) who showed that the heliocentric radial velocity
distribution of cluster members peaks at $22$~km s$^{-1}$
with a dispersion of $\sim3.5$~km s$^{-1}$. We adopt their
radial velocity membership selection criteria and retain
only those stars with 8~km s$^{-1}$~$<V_{rad}<$~36~km s$^{-1}$. This
leaves us with a final sample of 17 stars (see
Table~\ref{tabXX}) that will be used in the upcoming
analysis to estimate the distance to NGC~2264. 

Figure~\ref{figX} shows the proper motion vectors for
NGC~2264 stars. One notices that the proper motion
vectors point towards a common direction in the sky
suggesting the existence of a moving group structure
in NGC~2264 as observed in other nearby star-forming regions
(see e.g. Bertout \& Genova 2006; Galli et al.\ 2013).
Apparently the stellar proper motions in NGC~2264 are
not convergent enough to allow for an accurate convergent
point solution and membership analysis (which is clearly
beyond the scope of this paper) using the usual methods
and algorithms given in the literature (Jones 1971; de
Bruijne 1999; Galli et al.\ 2012), because of the large
proper motions errors involved and cluster concentration
(see Galli et al.\ 2012 for a more detailed discussion).
So we estimate an approximate location of the convergent
point using Eq.~(10) of de Bruijne\ (1999) that relates
the spatial velocity components with the convergent point
coordinates $(\alpha_{cp},\delta_{cp})$. 
We calculate the Galactic space motion vector for
the `classical' cluster members of NGC~2264 using the procedure
described in Johnson \& Soderblom (1987) with the proper motions,
radial velocities and parallaxes given in Table~\ref{tabX}.
HD~48055 is excluded from this analysis because it exhibits a
negative value for the radial velocity that is not consistent
with the remaining members (see discussion below). The mean
Galactic space motion vector with respect to the Sun is
$(U,V,W)=(-13.6,-17.3,-4.9)\pm(3.2,2.2,2.3)$~km/s. 
We find that the
mean space motion vector mentioned before is consistent
with a convergent point located at $(\alpha_{cp},\delta_{cp})
=(99\arcdeg,-22\arcdeg)\pm(6\arcdeg,10\arcdeg)$. 
The so-derived convergent point position should be interpreted
with care, because it does not take into account, e.g.,
the velocity dispersion of the cluster and the various
sources of errors discussed by Galli et al.\ (2012). However,
this strategy has proven to yield consistent results in
young nearby associations (see e.g. Mamajek 2005; Bertout \&
Genova 2006) and it will be used in this work only to provide
some guidance in our analysis. 

Once the convergent point position of a moving group has been
determined, it can be used to calculate kinematic parallaxes
of individual group members (see e.g. Bertout \& Genova 2006;
Galli et al.\ 2013). The individual kinematic parallaxes are
given by
\begin{equation}
\pi=\frac{A\,\mu_{\parallel}}{V_{rad}\tan\lambda}\, ,
\end{equation}
where $A=4.74047$~km~yr s$^{-1}$, $\lambda$ is the angular distance
between the convergent point and a given star in the group,
$V_{rad}$ is the radial velocity, and $\mu_{\parallel}$ is the
stellar proper motion component directed parallel to the great
circle that joins the star and the convergent point (see de
Bruijne 1999; Galli et al.\ 2012). This analysis is obviously
restricted to the 17~stars with known radial velocities (see
Table~\ref{tabXX}). We stress that the so-derived parallaxes
(i.e., distances)  should be seen as tentative values due to
the large proper motions errors given for a few cluster members
in our sample and the approximate location of the convergent
point derived from the spatial velocity of Hipparcos stars. In
order to make our parallax results less dependent on a single
combination of proper motion, radial velocity and convergent
point values we performed Monte Carlo simulations that sample
the uncertainties of these parameters. In each run we vary the
convergent point position within its error bars and assign a 
different value of proper motion and radial velocity for each 
star in the sample by resampling these quantities from a
Gaussian distribution where the mean and variance correspond,
respectively, to the measured value and its uncertainty. We
constructed a total of 1000 Monte Carlo realizations and present
our distance results in Figure~\ref{figXX}. We find a mean
distance of 340~pc  and a median distance of 300~pc. These
results are consistent with the mean distance derived from the
trigonometric parallaxes of the Hipparcos catalog (van
Leeuwen 2007) within the admittedly large error bars, and
with our distance estimate for the HH~124~IRS cluster. 

\subsection{Generalized kinematic distances to NGC~2264}

As a final exercise, we applied our generalized kinematic method to
the 17 stars of NGC~2264 listed  in Table \ref{tabXX}. The method 
failed on 4 of those stars because the sign of the proper motion 
component in the direction of Galactic longitude could not be reproduced 
for any distance. The mean and median distances obtained for the other 
13 stars are 453 and 391 pc, respectively. This is, again, in good
agreement with the Hipparcos and convergent point distances derived
earlier for NGC~2264 and with the kinematic distance of HH~124. As
a consequence, we favor a distance of order 400 pc both for HH~124~IRS
and for NGC~2264.

This new estimation of the distance is around two times lower than
values commonly used in the literature. The used kinematic methods
are biased to closer objects which proper motions are better determined.
The trigonometrical parallax, however, is a geometric method and does
not depend in any physical property of the star. Finally, NGC 2264
has an angular size of 250$'$, thus it is not unlikely that several
star forming regions are trough the line of sight at diferent
distances. 


\clearpage

\begin{figure*}
\epsscale{1}
\plotone{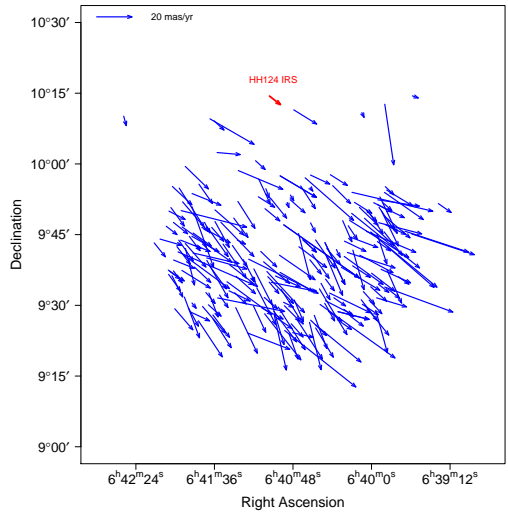}
\caption{Proper motions vectors for stars in NGC~2264. 
\label{figX}
}
\end{figure*}


\clearpage

\begin{figure*}
\epsscale{1}
\plotone{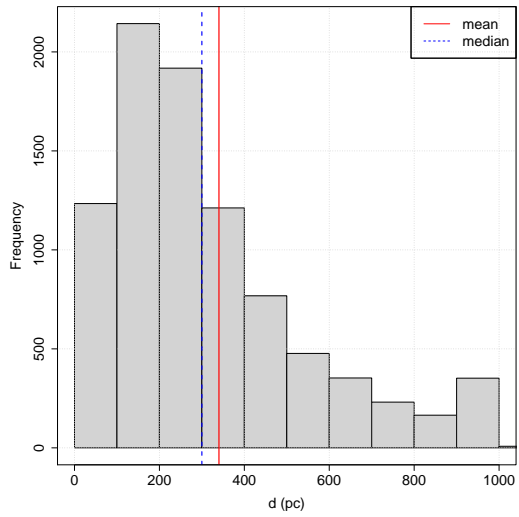}
\caption{Histogram of kinematic distances calculated with proper motions and radial velocities for the 17 stars given in Table~\ref{tabXX} after 1000 Monte Carlo realizations.  
\label{figXX}
}
\end{figure*}

}

\section{Conclusions}

We presented a series of deep VLA observations
of the compact cluster of radio sources around HH~124~IRS.
We analyze their spectral indices (between 4.8 and 7.5 GHz),
and variability. Also a rough estimate of proper motions was
performed to objects detected in both the Reipurth et al.\ (2002)
observations and those reported in this paper. The results
show that most of the sources detected by Reipurth et al.\ 
(2002) in the cluster are YSOs. We detected two additional
sources (both likely YSOs) in the compact cluster, bringing to 
at least 11 the total number of radio-bright YSOs in HH~124~IRS.
Eight of these sources are in an area
smaller than 1 arcmin$^2$. The importance of such compact 
clusters resides in that all of them can be observed in a single
pointing with most telescopes. Thanks to these characteristics, 
HH~124~IRS is an ideal region where to perform simultaneous
observations at different wavelengths to understand the 
variable radio emission and its correlation (or lack thereof)
with emission at other wavelengths.

{ A generalized kinematic method that takes into account both
the proper motions and the radial velocity was applied to the
sources in HH~124~IRS, and yields a distance of order 400 pc.
This is significantly smaller than the $\sim$800--900 pc distance 
usually assigned to the nearby open cluster NGC~2264 with 
which HH~124 is thought to be associated. However, a reanalysis 
of the Hipparcos parallaxes for members of NGC~2264, a convergent 
point approach applied to NGC~2264 members with known proper
motions, and a generalized kinematic analysis of sources in 
NGC~2264 all argue in favor of a distance of order 400 pc for 
NGC~2264 as well. Thus, we argue that both HH~124~IRS
and NGC~2264 are located about twice nearer than usually 
assumed.}


\acknowledgments
SAD, LFR and LL acknowledges the support
of DGAPA, UNAM, and of CONACyT (M\'exico).
This research has made use of the SIMBAD database, 
operated at CDS, Strasbourg, France.

\clearpage

\appendix

\section{Error introduced in the variability by pointing errors of the primary beam}

Using the standard pointing constants, the pointing error for the VLA antennas is generally 
10 to 20 arcseconds, and can be as bad as an arcminute (Rupen 1997).
To correct for this problem, a technique known as referenced pointing is used
to reduce this pointing error to a few arcsec. However, in observations below 
about 18 GHz (as in our case), referenced pointing is not generally used.

These pointing errors introduce an error in the flux density determination of the
sources and thus in the determination of possible variability. This error is
more serious for sources away from the pointing center, since there a small shift
may represent an important change in the primary beam response. In contrast, for sources
near the pointing center a small shift is not important, given the relatively flat
response of the primary beam there.

To include this error in our variability determinations, we start by noting
that the primary beam response can be approximated by a Gaussian function:

$$p(\theta) = \exp\Biggl(- 4~ \ln2~ {{\theta^2} \over {\theta^2_B}}\Biggr),$$

\noindent where $\theta$ is the displacement from the pointing center and
$\theta_B$ is the full width at half power of the primary beam.
In the radial direction the change in response with angular displacement is
given by

$${{\Delta p(\theta)}\over {\Delta \theta}} = 
\exp\Biggl(- 4~ \ln2~ {{\theta^2} \over {\theta^2_B}}\Biggr)
\Biggl(- 4~ \ln2~ {{1} \over {\theta^2_B}}\Biggr)~2 \theta,$$

\noindent where $\Delta \theta$ is the pointing error.
In this last equation we can see that the flux density error increases linearly with
the displacement from the pointing center. Finally, taking the absolute value of
the equation, we find that the relative error is given by

$${{\Delta p(\theta)}\over {p(\theta)}} =
\Biggl({{4~ \ln2} \over {\theta^2_B}}\Biggr)~2 \theta~\Delta \theta.$$

We have taken this effect approximately into account by adding in quadrature this
relative error to the other errors in the measurement, assuming $\Delta \theta =
30{''}$ as the typical value for the pointing error. 
For example, for a $\theta_B$ of $6'$, one expects errors
of order 8\% for a source at $\theta = 1'$ and of 23\% for a source at $\theta = 3'$.


\input{tabla1}

\input{table2}

\input{table3}


\clearpage

\begin{table*}
\small{
\begin{center}
\caption{Generalized distance determination to HH~124~IRS
\label{tab:kin}
\vspace{0.5cm}}
\begin{tabular}{lcccc}
\tableline\tableline
Rotation curve & $d_{V_{rad}}$ (pc) & $d_{\mu_\ell}$ (pc) & $d_{\mu_b}$ (pc) & $\bar{d}$ (pc)\\
\tableline
Blitz \& Rand &   620$^{+210}_{-200}$ &  430$^{+570}_{-170}$ &   280$^{+140}_{-70}$ &    356 $\pm$ 155 \\
Flat (IAU)      &   600$^{+200}_{-190}$ &  450$^{+450}_{-180}$ &   240$^{+180}_{-70}$ &    362 $\pm$ 148 \\
Reid et al.     &   550$^{+190}_{-170}$ &  390$^{+580}_{-150}$ &   240$^{+180}_{-70}$ &    336 $\pm$ 127 \\
Flat (Reid)    &   510$^{+170}_{-160}$ &  440$^{+560}_{-180}$ &   240$^{+190}_{-70}$ &    346 $\pm$ 114 \\
\tableline
\tablecomments{The four rotation curves considered are (i) Blitz \& Brand (1993), 
(ii) a flat curve with $\theta_0$ = 220 km s$^{-1}$, (iii) Reid et al.\ (2009), and (iv) a
flat curve with $\theta_0$ = 254 km s$^{-1}$ (see text). The four quoted distances
are those obtained (i) from a traditional kinematic method (based on radial velocities),
(ii) a kinematic method applied to the Galactic longitude component of the proper 
motion, (iii) a kinematic method applied to the Galactic latitude component of the proper 
motion, and (iv) the weighted mean of the previous three.}
\end{tabular}
\end{center}
}
\end{table*}


\clearpage

\begin{table*}
\small{
\begin{center}
\caption{Position (J2000), proper motion, radial velocity, and trigonometric parallax from the Hipparcos catalog (van Leeuwen~2007) for stars related to the NGC~2264 cluster.
\label{tabX}
\vspace{0.5cm}}
\begin{tabular}{lcccccccc}
\tableline\tableline
Star&$\alpha$&$\delta$&$\mu_{\alpha}\cos\delta$&$\mu_{\delta}$&Ref.&$V_{rad}$&Ref.&$\pi$\\
&(h:m:s)&$(^{\circ}\,\,\,\arcmin\,\,\,\arcsec)$&(mas/yr)&(mas/yr)&&(km/s)&&(mas)\\
\tableline

HD~47732                 	&	06	40	28.6	&	+09	49	04	&$	1.6	\pm	1.8	$&$	-4.5	\pm	1.7	$&	TYCHO2	&$	18.0	\pm	5.0	$&	1	&$	1.49	\pm	0.93	$\\
HD~47777                 	&	06	40	42.3	&	+09	39	21	&$	-0.6	\pm	1.4	$&$	-4.2	\pm	1.5	$&	UCAC4	&$	13.0	\pm	3.7	$&	2	&$	1.03	\pm	0.86	$\\
V*SMon                  	&	06	40	58.7	&	+09	53	45	&$	-0.7	\pm	1.0	$&$	-2.5	\pm	1.0	$&	UCAC4	&$	22.0	\pm	0.3	$&	3	&$	3.55	\pm	0.50	$\\
HD~47961                 	&	06	41	27.3	&	+09	51	14	&$	-1.3	\pm	1.4	$&$	-5.2	\pm	1.3	$&	TYCHO2	&$	23.4	\pm	1.8	$&	2	&$	2.22	\pm	0.69	$\\
HD~48055                 	&	06	41	49.7	&	+09	30	29	&$	-1.9	\pm	0.5	$&$	-3.3	\pm	0.7	$&	UCAC4	&$	-2.0	\pm	8.6	$&	4	&$	3.90	\pm	1.16	$\\
\tableline
HD~47662                 	&	06	40	13.4	&	+10	26	29	&$	-2.1	\pm	0.8	$&$	-1.4	\pm	0.8	$&	UCAC4	&$	-9.0	\pm	2.3	$&	4	&$	2.12	\pm	1.03	$\\
HD~47754                 	&	06	40	36.5	&	+10	23	31	&$	-2.1	\pm	0.5	$&$	-2.6	\pm	0.6	$&	UCAC4	&$	-3.0	\pm	2.2	$&	4	&$	1.98	\pm	1.14	$\\
HIP~32141                 	&	06	42	43.6	&	+08	51	19	&$	-3.1	\pm	0.9	$&$	-0.9	\pm	0.7	$&	UCAC4	&$	29.2	\pm	1.2	$&	5	&$	1.97	\pm	0.95	$\\
HIP~32245              	&	06	43	54.0	&	+09	03	50	&$	1.5	\pm	0.7	$&$	-2.4	\pm	0.6	$&	UCAC4	&$	---	$&	--	&$	3.94	\pm	0.95	$\\

\tableline
\end{tabular}
\tablecomments{The upper panel refers to the `classical' cluster members identified by Walker~(1956) and the lower panel refers to other Hipparcos stars in the cluster field (see Section~3).}
\tablerefs{
(1)~Ochsenbein~(1980); (2)~Kharchenko et al.~(2007); (3)~de Bruijne \& Eilers~(2012); (4)~Fehrenbach et al.~(1992); (5)~Gontcharov~(2006). 
}

\end{center}
}
\end{table*}


\clearpage

\begin{table*}
\scriptsize{
\begin{center}
\caption{Stars in NGC~2264 with known proper motion and radial velocity.
\label{tabXX}
\vspace{0.5cm}}
\begin{tabular}{lccccccc}
\tableline\tableline
Star&$\alpha$&$\delta$&$\mu_{\alpha}\cos\delta$&$\mu_{\delta}$&Ref.&$V_{rad}$&Ref.\\
&(h:m:s)&$(^{\circ}\,\,\,\arcmin\,\,\,\arcsec)$&(mas/yr)&(mas/yr)&&(km/s)\\
\tableline

Cl*NGC2264RMS1241	&	06	39	49.6	&	09	33	22	&$	-5.0	\pm	3.8	$&$	-5.6	\pm	3.8	$&	PPMXL	&$	15.5	\pm	1.5	$&	1	\\
Cl*NGC2264LBM677	&	06	39	56.4	&	09	43	32	&$	-6.0	\pm	3.8	$&$	-5.9	\pm	3.8	$&	PPMXL	&$	33.8	\pm	1.5	$&	1	\\
Cl*NGC2264LBM1207	&	06	40	01.4	&	09	31	06	&$	-5.8	\pm	3.8	$&$	-8.6	\pm	3.8	$&	PPMXL	&$	18.3	\pm	1.5	$&	1	\\
V*V594Mon	&	06	40	23.7	&	09	55	24	&$	-8.2	\pm	3.8	$&$	-11.4	\pm	3.8	$&	PPMXL	&$	25.0	\pm	1.5	$&	1	\\
V*NTMon	&	06	40	30.9	&	09	34	41	&$	-4.1	\pm	3.8	$&$	-10.4	\pm	3.8	$&	PPMXL	&$	17.0	\pm	1.5	$&	1	\\
HD261878	&	06	40	51.6	&	09	51	49	&$	-0.8	\pm	0.5	$&$	-3.2	\pm	0.6	$&	UCAC4	&$	15.0	\pm	10.0	$&	2	\\
Cl*NGC2264LBM5696	&	06	41	01.1	&	09	34	52	&$	-20.6	\pm	3.8	$&$	-5.7	\pm	3.6	$&	PPMXL	&$	21.6	\pm	1.5	$&	1	\\
HD261938	&	06	41	01.9	&	09	52	48	&$	-7.5	\pm	2.0	$&$	-12.9	\pm	1.8	$&	TYCHO2	&$	16.0	\pm	10.0	$&	2	\\
V*OQMon	&	06	41	06.9	&	09	29	24	&$	-9.6	\pm	3.8	$&$	-6.9	\pm	3.8	$&	PPMXL	&$	18.6	\pm	1.5	$&	1	\\
V*V817Mon	&	06	41	09.2	&	09	53	02	&$	-7.5	\pm	5.1	$&$	-6.6	\pm	5.1	$&	PPMXL	&$	19.7	\pm	1.5	$&	1	\\
Cl*NGC2264RMS3656	&	06	41	10.9	&	10	00	41	&$	-4.4	\pm	1.5	$&$	-5.6	\pm	1.5	$&	PPMXL	&$	18.2	\pm	1.5	$&	1	\\
V*V368Mon	&	06	41	18.4	&	09	39	41	&$	-5.4	\pm	3.8	$&$	-14.2	\pm	3.8	$&	PPMXL	&$	21.4	\pm	1.5	$&	1	\\
Cl*NGC2264FMS502	&	06	41	41.2	&	09	32	43	&$	-4.2	\pm	3.8	$&$	-6.9	\pm	3.8	$&	PPMXL	&$	32.1	\pm	1.5	$&	1	\\
Cl*NGC2264LBM8851	&	06	41	48.1	&	09	42	43	&$	-7.1	\pm	3.8	$&$	-3.9	\pm	3.8	$&	PPMXL	&$	17.8	\pm	1.5	$&	1	\\
Cl*NGC2264DS489	&	06	41	49.0	&	09	41	06	&$	-21.0	\pm	3.8	$&$	-6.7	\pm	3.8	$&	PPMXL	&$	16.4	\pm	1.5	$&	1	\\
Cl*NGC2264LBM9866	&	06	41	56.4	&	09	50	06	&$	-30.0	\pm	1.0	$&$	-10.5	\pm	1.0	$&	PPMXL	&$	16.1	\pm	1.5	$&	1	\\
Cl*NGC2264RMS5374	&	06	42	09.2	&	09	44	03	&$	-9.8	\pm	3.8	$&$	-3.9	\pm	3.8	$&	PPMXL	&$	17.2	\pm	1.5	$&	1	\\

\tableline
\end{tabular}
\tablecomments{We adopt an error of 1.5~km/s on the radial velocities given by F{\H u}r{\'e}sz et al.(2006) (see Section~2.3 of their paper).}
\tablerefs{(1)~F{\H u}r{\'e}sz et al.~(2006); (2)~Wilson~(1953)}
\end{center}
}
\end{table*}

\end{document}

%% file: tabla1.tex
\begin{deluxetable}{ccrcrcc}
\tabletypesize{\scriptsize}
\tablewidth{0pt}
\tablecolumns{7}
\tablecaption{Radio Sources Detected in the HH~124 Region. \label{tab:Rsour}}
\tablehead{ Name & \multicolumn{4}{c}{Flux Properties}	&Spectral& Source\\
\colhead{HH124-VLA } & \colhead{$f_{4.8}$(mJy)} &\colhead{Var.\,(\%)} & \colhead{$f_{7.5}$(mJy)} & \colhead{Var.\,(\%)}&\colhead{Index}& \colhead{type\tablenotemark{a}}\\
}\startdata
J064032.32+101644.0&(2.04$\pm$0.07$\pm$0.10)e+00&32.3$\pm$9.6&--&--&--&E?\\
J064032.44+101644.9&(1.16$\pm$0.07$\pm$0.06)e+00&30.5$\pm$10.8&--&--&--&E?\\
J064041.61+102036.6&(4.07$\pm$0.04$\pm$0.20)e+00&Extended&--&--&--&E\\
J064047.85+100829.3&(2.76$\pm$0.51$\pm$0.14)e-01&24.3$\pm$29.1&--&--&--&E?\\
J064049.82+101027.4&(9.20$\pm$1.70$\pm$0.46)e-02&$>$44.9$\pm$15.0&--&--&--&E?\\
J064050.28+101417.8&(5.22$\pm$0.74$\pm$0.26)e-02&41.6$\pm$18.8&(2.75$\pm$0.55$\pm$0.14)e-02&$>$19.3$\pm$29.4&-1.4$\pm$0.56&E?\\
J064050.63+101420.8&(3.78$\pm$0.46$\pm$0.19)e-02&20.1$\pm$18.7&$<$0.02&--&$<$-1.59$\pm$0.31&E?\\
J064053.63+101343.8&(3.45$\pm$0.32$\pm$0.17)e-02&34.0$\pm$16.0&(3.66$\pm$0.65$\pm$0.18)e-02&$>$46.0$\pm$20.2&0.13$\pm$0.47&E?\\
J064056.62+101408.7&(2.74$\pm$0.23$\pm$0.14)e-01&21.2$\pm$12.0&(1.84$\pm$0.16$\pm$0.09)e-01&39.6$\pm$13.1&-0.87$\pm$0.31&E?\\
J064058.51+101445.2&(3.24$\pm$0.30$\pm$0.16)e-02&35.3$\pm$12.7&(3.13$\pm$0.31$\pm$0.16)e-02&40.9$\pm$13.1&-0.08$\pm$0.34&E?\\
J064059.25+101756.8&(5.69$\pm$0.62$\pm$0.28)e-02&11.6$\pm$20.5&(3.62$\pm$0.50$\pm$0.18)e-02&32.1$\pm$17.2&-0.99$\pm$0.42&E?\\
J064059.64+101758.9&(2.69$\pm$0.46$\pm$0.13)e-02&$>$29.9$\pm$10.1&$<$0.02&--&$<$-0.99$\pm$0.41&E?\\
J064059.89+101214.4&(4.68$\pm$0.52$\pm$0.23)e-02&$>$55.1$\pm$7.6&(3.57$\pm$0.61$\pm$0.18)e-02&--\tablenotemark{b}&-0.59$\pm$0.47&E?\\
J064100.01+102222.7&(1.19$\pm$0.07$\pm$0.06)e+00&21.1$\pm$10.0&--&--&--&E?\\
J064101.12+101208.4&(8.15$\pm$0.60$\pm$0.41)e-02&14.8$\pm$14.7&(5.35$\pm$0.55$\pm$0.27)e-02&29.0$\pm$17.5&-0.92$\pm$0.32&E?\\
J064101.21+101450.1&(9.89$\pm$0.67$\pm$0.49)e-02&7.5$\pm$9.2&(7.81$\pm$0.54$\pm$0.39)e-02&27.4$\pm$10.8&-0.52$\pm$0.26&YSO?\\
J064101.22+101300.5&(4.14$\pm$0.43$\pm$0.21)e-02&15.2$\pm$20.7&(2.77$\pm$0.29$\pm$0.14)e-02&47.6$\pm$19.7&-0.88$\pm$0.36&E?\\
J064102.59+101503.5&(3.87$\pm$0.49$\pm$0.19)e-02&38.0$\pm$19.3&(3.83$\pm$0.32$\pm$0.19)e-02&29.0$\pm$14.0&-0.02$\pm$0.37&YSO\\
J064102.65+101501.8&(1.50$\pm$0.51$\pm$0.07)e-02&--\tablenotemark{b}&(2.67$\pm$0.27$\pm$0.13)e-02&33.5$\pm$15.6&1.26$\pm$0.79&YSO\\
J064103.39+101510.3&$<$0.01&--&(2.67$\pm$0.27$\pm$0.13)e-02&47.4$\pm$20.9&$>$1.87$\pm$0.39&YSO?\\
J064103.52+101506.2&(7.78$\pm$0.47$\pm$0.39)e-02&18.6$\pm$10.2&(8.32$\pm$0.59$\pm$0.42)e-02&15.2$\pm$12.8&0.15$\pm$0.26&YSO?\\
J064104.33+101240.9&(7.27$\pm$0.48$\pm$0.36)e-02&32.6$\pm$10.2&(7.87$\pm$0.60$\pm$0.39)e-02&13.8$\pm$16.3&0.17$\pm$0.27&E\\
J064104.45+101452.5\tablenotemark{c}&(3.41$\pm$0.36$\pm$0.17)e-02&51.3$\pm$12.1&(4.47$\pm$0.41$\pm$0.22)e-02&24.3$\pm$15.7&0.59$\pm$0.34&YSO?\\
J064104.58+101454.6&$<$0.01&--&(4.47$\pm$0.41$\pm$0.22)e-02&33.7$\pm$23.7&$>$1.26$\pm$0.35&YSO?\\
J064104.92+101340.0&(2.62$\pm$0.47$\pm$0.13)e-02&$>$44.1$\pm$9.7&(2.14$\pm$0.68$\pm$0.11)e-02&--\tablenotemark{b}&-0.44$\pm$0.81&E?\\
J064105.92+101551.7\tablenotemark{c}&(4.15$\pm$0.36$\pm$0.21)e-02&40.7$\pm$10.6&(3.35$\pm$0.37$\pm$0.17)e-02&28.3$\pm$16.4&-0.47$\pm$0.34&YSO?\\
J064106.53+101739.7&(4.50$\pm$0.41$\pm$0.22)e-02&27.0$\pm$17.9&(3.61$\pm$0.85$\pm$0.18)e-02&$>$36.6$\pm$17.6&-0.48$\pm$0.57&E?\\
J064106.77+100938.8&(3.09$\pm$0.03$\pm$0.15)e+00&Extended&--&--&--&E\\
J064107.90+102201.0&(6.47$\pm$0.50$\pm$0.32)e-01&14.1$\pm$11.7&--&--&--&E?\\
J064109.69+101455.7&(5.47$\pm$0.66$\pm$0.27)e-02&31.4$\pm$17.0&(5.40$\pm$1.30$\pm$0.27)e-02&$>$58.1$\pm$19.6&-0.03$\pm$0.61&E?\\
J064110.65+101538.2&(3.08$\pm$0.15$\pm$0.15)e-01&9.3$\pm$7.7&(3.89$\pm$0.05$\pm$0.19)e-01&6.3$\pm$3.1&0.51$\pm$0.19&YSO?\\
J064111.69+101417.6&(2.49$\pm$0.48$\pm$0.12)e-02&16.7$\pm$21.9&(2.16$\pm$0.46$\pm$0.11)e-02&--\tablenotemark{b}&-0.31$\pm$0.65&E?\\
J064112.18+101010.3\tablenotemark{c}&(2.74$\pm$0.04$\pm$0.14)e+01&12.3$\pm$2.3&--&--&--&E?\\
J064112.23+101417.2\tablenotemark{c}&(9.16$\pm$0.07$\pm$0.46)e-01&14.1$\pm$1.5&(8.75$\pm$0.06$\pm$0.44)e-01&10.0$\pm$3.0&-0.1$\pm$0.16&YSO?\\
J064112.49+101416.4&(1.18$\pm$0.10$\pm$0.06)e-01&17.5$\pm$14.5&(9.17$\pm$0.84$\pm$0.46)e-02&12.8$\pm$22.2&-0.55$\pm$0.31&E?\\
J064112.78+101219.1&(9.52$\pm$0.19$\pm$0.48)e-01&Extended&(4.83$\pm$0.21$\pm$0.24)e-01&Extended&--&E\\
J064114.16+101158.5&(2.61$\pm$0.01$\pm$0.13)e+01&Extended&(1.57$\pm$0.01$\pm$0.08)e+01&Extended&--&E\\
J064115.05+101148.8&(8.67$\pm$0.96$\pm$0.43)e-02&29.4$\pm$18.1&(3.62$\pm$0.70$\pm$0.18)e-02&--\tablenotemark{b}&-1.91$\pm$0.51&E?\\
J064115.18+101655.8&(2.87$\pm$0.38$\pm$0.14)e-02&$>$43.5$\pm$15.3&$<$0.02&--&$<$-0.34$\pm$0.33&E?\\
J064115.29+101126.3&(1.23$\pm$0.13$\pm$0.06)e-01&46.8$\pm$12.9&(1.01$\pm$0.18$\pm$0.05)e-01&--\tablenotemark{b}&-0.43$\pm$0.48&E?\\
J064116.18+101441.4&(2.28$\pm$0.15$\pm$0.11)e-01&13.0$\pm$12.3&(1.31$\pm$0.10$\pm$0.07)e-01&39.5$\pm$17.9&-1.21$\pm$0.27&E?\\
J064117.55+101533.4&(6.76$\pm$0.69$\pm$0.34)e-02&15.5$\pm$18.4&(4.99$\pm$0.71$\pm$0.25)e-02&27.9$\pm$27.8&-0.67$\pm$0.41&E?\\
J064118.28+101745.7&(4.13$\pm$0.64$\pm$0.21)e-02&$>$39.1$\pm$17.9&$<$0.06&--&$<$0.82$\pm$0.37&Star\\
J064119.82+101537.6&(5.38$\pm$0.76$\pm$0.27)e-02&$>$49.3$\pm$12.8&$<$0.04&--&$<$-0.71$\pm$0.35&E?\\
J064120.73+101323.4&(8.20$\pm$0.79$\pm$0.41)e-02&30.6$\pm$15.8&$<$0.06&--&$<$-0.58$\pm$0.26&E?\\
J064122.32+101439.7&(5.71$\pm$0.55$\pm$0.29)e-02&7.4$\pm$24.1&$<$0.07&--&$<$0.28$\pm$0.26&E?\\
J064123.26+101036.6&(2.31$\pm$0.18$\pm$0.12)e+00&13.8$\pm$11.3&--&--&--&E?\\
J064123.64+101231.7&(6.77$\pm$0.74$\pm$0.34)e-02&--\tablenotemark{b}&--&--&--&E?\\
J064124.11+101949.7&(4.77$\pm$0.41$\pm$0.24)e-01&32.0$\pm$11.5&--&--&--&E?\\
J064126.64+101008.6&(3.38$\pm$0.26$\pm$0.17)e-01&27.0$\pm$17.2&--&--&--&E?\\

\enddata
\tablenotetext{a}{ E = Extragalactic, YSO = Young Stellar Object. Interrogation simbols
indicates that we do no discard an extragalactic or galactic origin.}
\tablenotetext{b}{Source not detected at three times the noise level on individual epochs,
but detected on the image of the concatenated epochs.}
\tablenotetext{c}{Time variable sources on a timescale of 3 days.}
\end{deluxetable}

%% file: table2.tex
\begin{table*}[htb]
\small
\begin{center}
\caption{Radio, infrared and optical counterparts of some radio-sources around HH 124IRS.}
\begin{tabular}{lccccc}\hline\hline
HH124-VLA Name                    &Radio& WISE & 2MASS & SDSS &   Origin     \\
\hline
J064041.61+102036.6 &     &  Y   &   Y   &   Y  & E\\
J064101.21+101450.1 &VLA 7&  N   &   N   &   N  &      YSO?\\
J064102.59+101503.6 &VLA 1&  Y   &   Y   &   N  &      YSO     \\
J064102.65+101501.8 &VLA 9&  Y   &   Y   &   N  &      YSO     \\
J064103.52+101506.2 &VLA 2 & N   &   N   &   N  &YSO?\\
J064104.45+101452.5 &VLA 10& N   &   N   &   N  &YSO?\\
J064110.65+101538.2 &VLA 4 & N   &   N   &   N  &YSO?\\
J064112.23+101417.2 &VLA 5 & N   &   N   &   N  &YSO?\\
J064114.16+101158.5 &VLA 6 & N   &   N   &   N  &E\\
J064104.33+101240.9 &&  Y   &   Y   &   Y  & E\\
J064106.77+100938.8 &&  Y   &   N   &   N  & E?\\
J064112.18+101010.3 &&  Y   &   N   &   N  & E?\\
J064115.29+101126.3 &&  Y   &   N   &   N  & E?\\
J064118.28+101745.7 &&  Y   &   Y   &   Y  &     Star     \\
J064120.73+101323.4 &&  Y   &   N   &   N  & E?\\
J064126.64+101008.6 &&  Y   &   N   &   N  & E?\\
\hline\hline
   \label{tab:counterparts}
   \end{tabular}
 \end{center}
\end{table*}

%% file: table3.tex
\begin{table*}[htb]
\small
\begin{center}
\caption{Proper motions of some radio-sources.}
\begin{tabular}{lcccc}\hline\hline
                                & $\mu_{\alpha}\cos{\delta}$  &   $\mu_{\delta}$  &   $\mu_{\rm total}$& \\
HH124-VLA Name                            & (mas yr$^{-1}$)  &  (mas yr$^{-1}$) &   (mas yr$^{-1}$)  & Type\\
\hline
J064056.62+101408.7         & $-5.00\pm2.68$  &   $-4.59\pm2.74$  &   $6.78\pm2.71$  &   E?\\
J064101.21+101450.1=VLA 7   & $-8.99\pm1.34$  &   $-6.01\pm1.38$  &   $10.81\pm1.35$ &   YSO?\\
J064102.59+101503.6=VLA 1   & $-2.85\pm2.16$  &   $-5.95\pm2.21$  &   $6.59\pm2.20$  &   YSO\\
J064102.65+101501.8=VLA 9   & $-4.54\pm1.27$  &   $-5.61\pm1.31$  &   $7.21\pm1.30$  &   YSO\\
J064103.52+101506.2=VLA 2   & $-1.26\pm1.52$  &   $-7.07\pm1.61$  &   $7.18\pm1.61$  &   YSO?\\
J064104.45+101452.5=VLA 10  & $-2.03\pm1.97$  &   $-6.59\pm1.99$  &   $6.89\pm1.99$  &   YSO?\\
J064110.65+101538.2=VLA 4   & $-8.09\pm0.67$  &   $-5.59\pm0.65$  &   $9.83\pm0.66$  &   YSO?\\
J064112.23+101417.2=VLA 5   & $-7.34\pm0.61$  &   $-1.17\pm0.62$  &   $7.43\pm0.61$  &   YSO?\\
J064114.16+101158.5=VLA 6   & $-0.99\pm2.54$  &   $-1.18\pm2.65$  &   $1.54\pm2.61$  &   E\\
\hline\hline
   \label{tab:pm}
   \end{tabular}
 \end{center}
\end{table*}

%% file: paper.bbl
\begin{thebibliography}{}

\bibitem[Adelman-McCarthy 
\& et al.(2011)]{2011yCat.2306....0A} Adelman-McCarthy, J.~K., \& et al.\ 2011, VizieR Online Data Catalog, 2306, 0 

\bibitem[Andre(1996)]{1996ASPC...93..273A} Andre, P.\ 1996, Radio Emission 
from the Stars and the Sun, 93, 273 

\bibitem[Andre et al.(1992)]{1992ApJ...401..667A} Andre, P., Deeney, B.~D., 
Phillips, R.~B., \& Lestrade, J.-F.\ 1992, \apj, 401, 667 

\bibitem[Anglada et al.(1998)]{1998AJ....116.2953A} Anglada, G., 
Villuendas, E., Estalella, R., et al.\ 1998, \aj, 116, 2953 

\bibitem[Baxter et al.(2009)]{2009AJ....138..963B} Baxter, E.~J., Covey, 
K.~R., Muench, A.~A., et al.\ 2009, \aj, 138, 963 

\bibitem[Bertout 
\& Genova(2006)]{Bertout(2006)} Bertout, C., \& Genova, F.\ 2006, \aap, 460, 499 

\bibitem[Brand 
\& Blitz(1993)]{1993A&A...275...67B} Brand, J., \& Blitz, L.\ 1993, \aap, 275, 67 

\bibitem[Binney 
\& Merrifield(1998)]{1998gaas.book.....B} Binney, J., \& Merrifield, M.\ 1998, Galactic astronomy / James Binney and Michael Merrifield.~ Princeton, NJ : Princeton University Press, 1998.~ (Princeton series in astrophysics) QB857 .B522 1998  (\$35.00),  

\bibitem[Cutri et al.(2003)]{2003tmc..book.....C} Cutri, R.~M., Skrutskie, 
M.~F., van Dyk, S., et al.\ 2003, ''The IRSA 2MASS All-Sky Point Source 
Catalog, NASA/IPAC Infrared Science Archive.~A href=''http://irsa.ipac.caltech.edu/applications/Gator/''

\bibitem[Cutri 
\& et al.(2012)]{2012yCat.2311....0C} Cutri, R.~M., \& et al.\ 2012, VizieR Online Data Catalog, 2311, 0 

\bibitem[Dahm 
\& Simon(2005)]{Dahm(2005)} Dahm, S.~E., \& Simon, T.\ 2005, \aj, 129, 829 

\bibitem[Dahm et al.(2007)]{Dahm(2007)} Dahm, S.~E., Simon, T., 
Proszkow, E.~M., \& Patten, B.~M.\ 2007, \aj, 134, 999 

\bibitem[de Bruijne(1999)]{deBruijne(1999)} de Bruijne, J.~H.~J.\ 1999, 
\mnras, 306, 381 

\bibitem[de Bruijne 
\& Eilers(2012)]{deBruijne(2012)} de Bruijne, J.~H.~J., \& Eilers, A.-C.\ 2012, \aap, 546, A61 

\bibitem[Dehnen \& Binney(1998)]{1998MNRAS.298..387D} Dehnen, W., \& Binney, J.~J.\ 1998, \mnras, 298, 387 

\bibitem[Dulk(1985)]{1985ARA&A..23..169D} Dulk, G.~A.\ 1985, \araa, 23, 169

\bibitem[Dzib et al.(2013)]{2013arXiv1307.5105D} Dzib, S.~A., Loinard, L., 
Mioduszewski, A.~J., et al.\ 2013, arXiv:1307.5105 

\bibitem[Fehrenbach et 
al.(1992)]{Fehrenbach(1992)} Fehrenbach, C., Burnage, R., \& Figuiere, J.\ 1992, \aaps, 95, 541 

\bibitem[Flaccomio et 
al.(1999)]{Flaccomio(1999)} Flaccomio, E., Micela, G., Sciortino, S., et al.\ 1999, \aap, 345, 521 

\bibitem[Flaccomio et 
al.(2000)]{Flaccomio(2000)} Flaccomio, E., Micela, G., Sciortino, S., et al.\ 2000, \aap, 355, 651 

\bibitem[Flaccomio et 
al.(2006)]{Flaccomio(2006)} Flaccomio, E., Micela, G., \& Sciortino, S.\ 2006, \aap, 455, 903 


\bibitem[F{\H u}r{\'e}sz et al.(2006)]{Furesz(2006)} F{\H 
u}r{\'e}sz, G., Hartmann, L.~W., Szentgyorgyi, A.~H., et al.\ 2006, \apj, 
648, 1090

\bibitem[Galli et 
al.(2012)]{Galli(2012)} Galli, P.~A.~B., Teixeira, R., Ducourant, C., Bertout, C., \& Benevides-Soares, P.\ 2012, \aap, 538, A23 

\bibitem[Galli et 
al.(2013)]{Galli(2013)} Galli, P.~A.~B., Bertout, C., Teixeira, R., \& Ducourant, C.\ 2013, \aap, 558, A77 

\bibitem[G{\'o}mez et al.(2008)]{2008ApJ...685..333G} G{\'o}mez, L., 
Rodr{\'{\i}}guez, L.~F., Loinard, L., et al.\ 2008, \apj, 685, 333 

\bibitem[Gontcharov(2006)]{Gontcharov(2006)} Gontcharov, G.~A.\ 2006, 
Astronomy Letters, 32, 759 

\bibitem[H{\o}g et 
al.(2000)]{Tycho2} H{\o}g, E., Fabricius, C., Makarov, V.~V., et al.\ 2000, \aap, 355, L27 

\bibitem[Kharchenko et al.(2007)]{Kharchenko(2007)} Kharchenko, N.~V., 
Scholz, R.-D., Piskunov, A.~E., R{\"o}ser, S., 
\& Schilbach, E.\ 2007, Astronomische Nachrichten, 328, 889 

\bibitem[Johnson 
\& Soderblom(1987)]{Johnson(1987)} Johnson, D.~R.~H., \& Soderblom, D.~R.\ 1987, \aj, 93, 864 

\bibitem[Jones(1971)]{Jones(1971)} Jones, D.~H.~P.\ 1971, \mnras, 
152, 231 

\bibitem[Lamm et 
al.(2004)]{Lamm(2004)} Lamm, M.~H., Bailer-Jones, C.~A.~L., Mundt, R., Herbst, W., \& Scholz, A.\ 2004, \aap, 417, 557 

\bibitem[Mamajek(2005)]{Mamajek(2005)} Mamajek, E.~E.\ 2005, \apj, 
634, 1385 

\bibitem[Margulis et al.(1988)]{1988ApJ...333..316M} Margulis, M., Lada, 
C.~J., \& Snell, R.~L.\ 1988, \apj, 333, 316 

\bibitem[Neri et 
al.(1993)]{Neri(1993)} Neri, L.~J., Chavarria-K., C., \& de Lara, E.\ 1993, \aaps, 102, 201

 \bibitem[Ochsenbein(1980)]{Ochsenbein(1980)} Ochsenbein, F.\ 1980, 
Bulletin d'Information du Centre de Donnees Stellaires, 19, 74 

\bibitem[Perez et al.(1987)]{Perez(1987)} Perez, M.~R., The, P.~S., 
\& Westerlund, B.~E.\ 1987, \pasp, 99, 1050 


\bibitem[Ram{\'{\i}}rez et al.(2004)]{Ramirez(2004)} Ram{\'{\i}}rez, 
S.~V., Rebull, L., Stauffer, J., et al.\ 2004, \aj, 127, 2659 

\bibitem[Rebull et al.(2002)]{Rebull(2002)} Rebull, L.~M., Makidon, 
R.~B., Strom, S.~E., et al.\ 2002, \aj, 123, 1528 

\bibitem[Reid et al.(2009)]{2009ApJ...700..137R} Reid, M.~J., Menten, 
K.~M., Zheng, X.~W., et al.\ 2009, \apj, 700, 137 

\bibitem[Reipurth et al.(1999)]{1999AJ....118..983R} Reipurth, B.,
Rodr{\'{\i}}guez, L.~F., \& Chini, R.\ 1999, \aj, 118, 983

\bibitem[Reipurth et al.(2002)]{Reipurth(2002)} Reipurth, B., 
Rodr{\'{\i}}guez, L.~F., Anglada, G., \& Bally, J.\ 2002, \aj, 124, 1045 

\bibitem[Reipurth et al.(2004)]{2004AJ....127.1736R} Reipurth, B.,
Rodr{\'{\i}}guez, L.~F., Anglada, G., \& Bally, J.\ 2004, \aj, 127, 1736


\bibitem[Rodriguez
\& Reipurth(1994)]{1994A&A...281..882R} Rodriguez, L.~F., \& Reipurth, B.\ 1994, \aap, 281, 882

\bibitem[Rodr{\'{\i}}guez
\& Reipurth(1996)]{1996RMxAA..32...27R} Rodr{\'{\i}}guez, L.~F., \& Reipurth, B.\ 1996, \rmxaa, 32, 27


\bibitem[Rodr{\'{\i}}guez et al.(2000)]{2000AJ....119..882R}
Rodr{\'{\i}}guez, L.~F., Delgado-Arellano, V.~G., G{\'o}mez, Y., et al.\
2000, \aj, 119, 882


\bibitem[Rodr{\'{\i}}guez(1994)]{1994RMxAA..29...69R} Rodr{\'{\i}}guez, 
L.~F.\ 1994, \rmxaa, 29, 69

\bibitem[Rodriguez(1997)]{1997IAUS..182...83R} Rodriguez, L.~F.\ 1997, 
Herbig-Haro Flows and the Birth of Stars, 182, 83 


\bibitem[Rodr{\'{\i}}guez 
\& Reipurth(1998)]{1998RMxAA..34...13R} Rodr{\'{\i}}guez, L.~F., \& Reipurth, B.\ 1998, \rmxaa, 34, 13 

\bibitem[Rodr{\'{\i}}guez et al.(2012)]{2012RMxAA..48..243R} 
Rodr{\'{\i}}guez, L.~F., Dzib, S.~A., Loinard, L., et al.\ 2012, \rmxaa, 
48, 243

\bibitem[Roeser et al.(2010)]{PPMXL} Roeser, S., Demleitner, 
M., \& Schilbach, E.\ 2010, \aj, 139, 2440 

\bibitem[Rupen 1997]{} Rupen, M. P. 1997, VLA Test Memorandum No. 202: 
Referenced Pointing at the VLA (www.vla.nrao.edu/memos/test/202/)



\bibitem[Sagar 
\& Joshi(1983)]{Sagar(1983)} Sagar, R., \& Joshi, U.~C.\ 1983, \mnras, 205, 747 

\bibitem[Sch{\"o}nrich et al.(2010)]{2010MNRAS.403.1829S} Sch{\"o}nrich, 
R., Binney, J., \& Dehnen, W.\ 2010, \mnras, 403, 1829 

\bibitem[Sung et al.(1997)]{Sung(1997)} Sung, H., Bessell, M.~S., 
\& Lee, S.-W.\ 1997, \aj, 114, 2644 


\bibitem[Sung et al.(2004)]{Sung(2004)} Sung, H., Bessell, M.~S., 
\& Chun, M.-Y.\ 2004, \aj, 128, 1684 


\bibitem[van Leeuwen(2007)]{vanLeeuwen(2007)} van Leeuwen, F.\ 2007, 
Astrophysics and Space Science Library, 350,  

\bibitem[van Leeuwen(2009)]{vanLeeuwen(2009)} van Leeuwen, F.\ 2009, \aap, 497, 209 

\bibitem[Walker(1956)]{Walker(1956)} Walker, M.~F.\ 1956, \apjs, 2, 
365 

\bibitem[Wilson(1953)]{Wilson(1953)} Wilson, R.~E.\ 1953, Carnegie 
Institute Washington D.C.~Publication, 0 


\bibitem[Wu et al.(2009)]{Wu(2009)} Wu, Z.-Y., Zhou, X., Ma, J., 
\& Du, C.-H.\ 2009, \mnras, 399, 2146 

\bibitem[Zacharias et al.(2013)]{UCAC4} Zacharias, N., Finch, 
C.~T., Girard, T.~M., et al.\ 2013, \aj, 145, 44 


\end{thebibliography}
